\documentclass[twocolumn,pra,showpacs,superscriptaddress]{revtex4}

\usepackage{graphicx}
\usepackage{dcolumn}
\usepackage{ulem}
\usepackage{slashed}

\begin{document}

\title{Pump-probe nonlinear magneto-optical rotation with frequency modulated light}

\author{S.~Pustelny}
\affiliation{Instytut Fizyki im.\ M. Smoluchowskiego, Uniwersytet
Jagiello\'{n}ski, Reymonta 4, 30-059 Krak\'{o}w, Poland}
\author{D.~F.~Jackson Kimball}
\affiliation{Department of Physics, California State University --
East Bay, 25800 Carlos Bee Blvd., Hayward, CA 94542, USA}
\author{S.~M.~Rochester}
\affiliation{Department of Physics, University of California at
Berkeley, Berkeley, CA 94720-7300, USA}
\author{V.~V.~Yashchuk}
\affiliation{Advanced Light Source Division, Lawrence Berkeley
National Laboratory, Berkeley CA 94720, USA}
\author{W.~Gawlik}
\affiliation{Instytut Fizyki im.\ M. Smoluchowskiego, Uniwersytet
Jagiello\'{n}ski, Reymonta 4, 30-059 Krak\'{o}w, Poland}
\author{D.~Budker}
\affiliation{Department of Physics, University of California at
Berkeley, Berkeley, CA 94720-7300, USA} \affiliation{Nuclear Science
Division, Lawrence Berkeley National Laboratory, Berkeley CA 94720,
USA}

\date{\today}

\begin{abstract}
Specific types of atomic coherences between Zeeman sublevels can be
generated and detected using a method based on nonlinear
magneto-optical rotation with frequency modulated light. Linearly
polarized, frequency modulated light is employed to selectively
generate ground-state coherences between Zeeman sublevels for which
$\Delta m=2$ and $\Delta m=4$ in $^{85}$Rb and $^{87}$Rb atoms, and
additionally $\Delta m=6$ in $^{85}$Rb. The atomic coherences are
detected with a separate, unmodulated probe light beam. Separation
of the pump and probe beams enables independent investigation of the
processes of creation and detection of the atomic coherences. With
the present technique the transfer of the Zeeman coherences,
including high-order coherences, from excited to ground state by
spontaneous emission has been observed.
\end{abstract}

\maketitle

\section{Introduction}

The use of light to create and detect coherences between atomic
states, and the evolution of those coherences when the atomic states
are nondegenerate, is a powerful method for precision spectroscopy,
extensively employed for measurements of external fields, tests of
fundamental symmetries (see, for example, reviews
\cite{AlexandrovReview} and \cite{RevModPhys}), and and frequency
standards \cite{Vanier05apb}. In this work, we report an
experimental technique to selectively create and detect atomic
coherences between Zeeman sublevels ($\Delta m =2,4,6$) in the
ground states of alkali atoms. We employ two independent laser
beams, one for creation of the ground state coherences and a second
for detection of the coherences, allowing detailed experimental
studies of the optical pumping and probing processes.  Selective
creation and detection of the $\Delta m = 6$ coherence was achieved
for the first time, and transfer of high-order Zeeman coherences
from the excited state to the ground state via spontaneous emission
was observed.

Resonant nonlinear magneto-optical rotation (NMOR) has been studied
for over 30 years (see the review \cite{RevModPhys} and references
therein). The essence of the effect is light-intensity-dependent
rotation of the polarization plane of a beam of light propagating
through a medium placed in a magnetic field. Most often the effect
is studied in the Faraday geometry, in which the magnetic field is
applied along the direction of light propagation. The NMOR signals
can either be observed as a function of magnetic field with the
laser frequency fixed (magnetic-field domain), or as a function of
the laser frequency with the magnetic field fixed (spectral domain).
There are several effects that contribute to NMOR signals. These
contributions can be distinguished in the magnetic-field domain as
they typically give rise to optical rotation that peaks at different
values of the magnetic field (leading to a group of nested
dispersive features). In general, the width of the narrowest feature
is determined by the effective rate of relaxation of atomic
ground-state coherences. In evacuated (buffer-gas-free) cells with
anti-relaxation coating, ground-state coherences can be preserved
for 500 ms or longer, leading to NMOR signals with sub-Hertz widths
\cite{BudkerRamsey,BudkerAntirelaxation}.

The dispersive features in the magnetic field dependence observed in
a NMOR experiment with unmodulated light are centered at zero
magnetic field. This restricts magnetic field measurements using
NMOR to values of the magnetic field for which Larmor frequencies
are smaller than the ground-state relaxation rate. However,
combining the NMOR effect with the technique of synchronous optical
pumping \cite{BellBloom} allows one to obtain similarly narrow NMOR
features centered at non-zero magnetic fields, including the
Earth-field range important for many applications (see Ref.\
\cite{AlexandrovReview} and references therein). Such resonances
generally appear when a harmonic of the modulation frequency is a
particular multiple of the Larmor frequency.

The synchronous pumping of atoms in NMOR enables a selective
creation of high-rank atomic polarization moments (PMs) associated
with high-order Zeeman coherences. The PMs are described by the
density matrix elements in the irreducible tensorial basis
\cite{BudkerSelective,Ale93} (see Section \ref{sec:metodology}
below). Such moments and associated atomic coherences have drawn
attention \cite{Lobodzinski96,Suter93,Matsko03} because they may
enhance nonlinear optical effects important in such applications as
quantum gates \cite{Turchette95}, electromagnetically induced
transparency \cite{Harris97,Arimondo96}, and magnetometry
\cite{BudkerSelective,Alexandrov97,Okunevich01,Stahler01}.

In this paper we investigate selective creation and detection of
atomic ground-state PMs of rank $\kappa=2$, $\kappa=4$, and
$\kappa=6$, corresponding to coherences between Zeeman sublevels
with $\Delta m = 2,4$ and 6, respectively, via interaction with
linearly polarized, frequency modulated light (FM NMOR)
\footnote{Magneto-optical rotation in pump-probe arrangement without
frequency modulation has been studied previously (see Section 8c in
Ref.\ \cite{RevModPhys} and references therein).}. Two independent
light beams interacting with atoms via different excited states
enable detailed analysis of the processes of creation and detection
of atomic multipoles in the ground state. Specifically, we have
measured the signal dependences on pump and probe-beam tuning and
intensity. We have observed that the signals depend differently on
pump- and probe-light parameters, which is important for
optimization of magnetometry schemes based on FM NMOR. Separated
pump and probe beams may also be necessary for a more efficient
method of production of high-rank multipoles, wherein pumping and
probing are done at different harmonics of the Larmor frequency
\cite{HexAt2Larmor}.

The article is organized as follows. In Section \ref{sec:metodology}
the FM NMOR method is described in detail. The symmetries of the
atomic PMs are shown and mechanism by which a specific PM is
generated is explained. In Section \ref{sec:setup} the experimental
setup for pump-probe FM NMOR is discussed. Section \ref{sec:results}
presents the experimental results and their analysis. In particular,
the dependences of the FM NMOR-resonance amplitudes and widths on
pump and probe intensities and frequencies are presented and
discussed. Also the results are shown demonstrating coherences
transfer via spontaneous emission. Finally, conclusions are
summarized in Section \ref{sec:conclusion}.

\section{Generation and detection of atomic coherences and atomic
polarization moments} \label{sec:metodology}

The density matrix written in the $M,M'$ representation for a state
with total angular momentum $F$ can be decomposed into PMs of rank
$\kappa=0,1,\ldots,2F$ (also known as multipoles of order
$k=2^\kappa$) with components $q=-\kappa,-\kappa+1,\ldots,\kappa$:
\begin{equation}\label{eq:PMdecomposition}
    \rho^{(\kappa)}_q=\sum^F_{M,M'=-F}(-1)^{F-M'}\langle
    F,M,F,-M'|\kappa,q\rangle\rho_{MM'},
\end{equation}
where $\langle\ldots|\ldots\rangle$ are the Clebsch-Gordan
coefficients. The PM components transform as
\begin{equation}\label{eq:PMrotation}
    \rho^{(\kappa)}_q\rightarrow e^{-iq\phi}\rho^{(\kappa)}_q
\end{equation}
when the PM rotates by an angle $\phi$ about the quantization axis.
Interaction of light with an unpolarized sample (having only the
$\rho^{(0)}_0$ moment, which is proportional to the population
distribution) causes redistribution of atomic population and
creation of coherences among the Zeeman sublevels, generating
higher-order moments (components $\rho^{(\kappa)}_q$ are related to
coherences with $\Delta M=|q|$). With a single photon-atom
interaction, a PM of rank $\kappa\leq 2$ with components $|q|\leq 2$
can be generated. Creation of PMs with higher components $|q|>2$
requires multi-photon interactions between light and matter. Once
the PMs are created in the medium, the same number of photon
interactions as needed for their creation are required for the PMs
to affect the properties of the transmitted light. This can be
understood by recalling that a photon is a spin-one particle, so it
is described by a tensor operator of rank $\kappa\geq 2$.

Generally if the light intensity is sufficiently high to produce the
high-rank PM, it will also produce all the lower-rank moments with
greater or equal efficiency. Therefore it has been difficult to
distinguish effects related to high-rank PMs from those associated
with lower-rank PMs.

One method for generating and detecting specific PMs in a medium, as
reported in Ref.\ \cite{BudkerSelective}, is FM NMOR, in which
pump-light is modulated periodically bringing the light frequency on
and off resonance with the atoms, and effectively creating optical
pumping pulses. With linearly polarized light, atomic PMs of even
rank \footnote{At high light power, odd rank PMs can be generated
with a magnetic field present \cite{Bud2000AOC}.} are created, which
precess around the magnetic field, i.e., undergo continuous rotation
according to Eq.\ (\ref{eq:PMrotation}), with $\phi=\Omega_Lt$,
where $\Omega_L$ is the Larmor frequency. If the modulation
frequency $\Omega_{mod}$ and the Larmor frequency are not
commensurate, the transverse polarization (components with $q\ne0$)
created in successive pumping cycles will add with different phases,
and will wash out after multiple pumping pulses. However, if
$\Omega_{mod}=n\Omega_L$, where $n$ is an even number, the
transverse PMs with less than $n$-fold symmetry ($0<|q|<n$) will
average out, but those with $|q|\ge n$ will be reinforced. This can
be visualized using a method described, for example, in Ref.\
\cite{Visualization}.
\begin{figure}
    \includegraphics{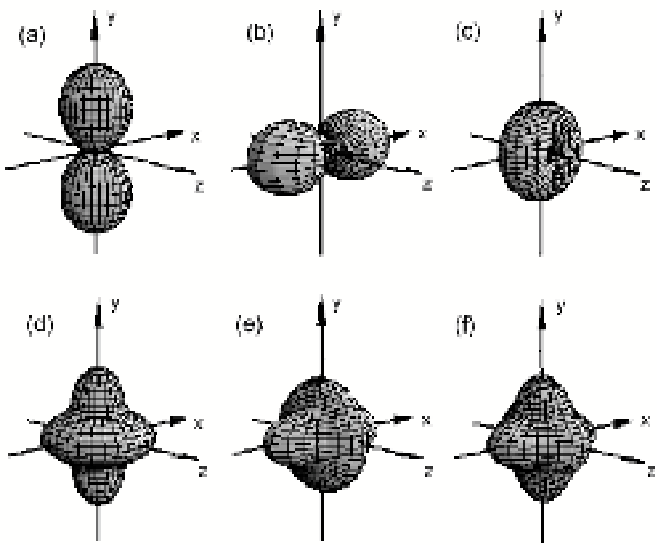}
    \caption{Visualization of the averaging of the PMs. The surface
    distance from the origin in a given direction is equal to the
    probability of finding the highest projection of the angular
    momentum ($M=F$) in that direction. The
    quadrupole and hexadecapole moments created in the medium due to
    interaction with light polarized along $y$-axis are shown in (a)
    and (d). After creation the atomic polarization rotates around
    the magnetic field applied along $z$-axis. If the pumping is
    modulated at four times the Larmor frequency newly created
    multipoles (a) and (d) are rotated by $90^\circ$ with respect to
    the previously created quadrupole (b) and hexadecapole (e)
    moments. The resultant polarization of the medium after many
    pumping cycles related to quadrupole and hexadecapole is shown
    in (c) and (f). Since the averaged quadrupole is symmetric with
    respect to rotation around $z$-axis the time-dependent changes
    of the optical anisotropy of the medium affecting light
    propagating along $z$-axis are related only to the hexadecapole
    moment. The plots are drawn for an $F=2$ state.} \label{fig:PM}
\end{figure}
On a particular cycle of the optical-pumping modulation (thought of
as a short pulse for the purposes of this illustration), PMs of all
even ranks up to $\kappa=2F$ are created in the ground state, with
the lowest-order moment (quadrupole) dominating at low power. The
quadrupole moment [Fig.\ \ref{fig:PM}(a)] has an axis along the
light polarization ($y$-axis) and a corresponding symmetry of order
2 about the $z$-axis. If modulation occurs at $4\Omega_{L}$, the PM
will have rotated one-fourth of a cycle ($90^\circ$) by the next
pump pulse [Fig.\ \ref{fig:PM}(b)]. The average of the rotated PM
and the newly pumped quadrupole moment is symmetric with respect to
the $z$-axis, i.e., the transverse polarization has averaged out
[Fig.\ \ref{fig:PM}(c)]. (Any remaining transverse polarization
would have fourth-fold symmetry with respect to the $z$-axis, which
cannot be supported by the quadrupole moment, with
$|q|\le\kappa=2$.) The longitudinal quadrupole component
$\rho^{(2)}_0$, symmetric with respect to the $z$-axis, does remain,
but it is not affected by the magnetic field nor does it cause
optical rotation. Because the transverse quadrupole moment has been
destroyed, the transverse hexadecapole moment is now the
lowest-order contributor to the FM NMOR signal (for $F=2$ it is the
only transverse moment remaining). For this moment, the pumped PM
[Fig.\ \ref{fig:PM}(d)] averaged with the PM rotated by $90^\circ$
[Fig.\ \ref{fig:PM}(e)] does leave transverse polarization remaining
[Fig.\ \ref{fig:PM}(f)], because the hexadecapole moment can support
fourth-fold symmetries. (The second-fold symmetry, corresponding to
$|q|=\Delta M=2$ is destroyed, but the $\rho^{(4)}_{\pm4}$
components remain.) Thus the low-order transverse PMs are
eliminated, leaving, in general, only PMs with $|q|$ greater than or
equal to $n$.

\begin{figure}
    \includegraphics{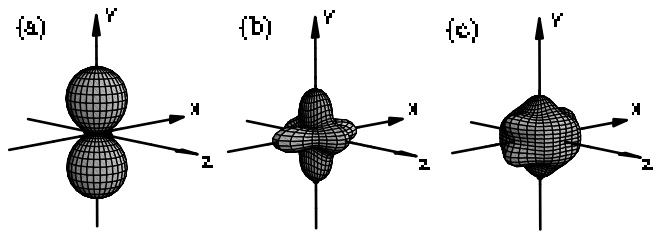}
    \caption{The averaged PMs: two-fold symmetric quadrupole (a),
    four-fold symmetric hexadecapole (b), and six-fold symmetric
    hexacontatetrapole (c). Each of the averaged PMs is associated
    with the Zeeman coherences in the ground state: averaged
    quadrupole with $\Delta m=2$, averaged hexadecapole with $\Delta
    m=4$, and averaged hexacontatetrapole with $\Delta m=6$. The
    plots are drawn for an F=3 state.} \label{fig:AveragedPMs}
\end{figure}

The precession of the transverse PMs in the magnetic field results
in time-varying optical rotation in the probe light beam. In our
experimental arrangement, the signal is proportional to the
amplitude of the optical rotation at various harmonics of the
modulation frequency. The optical pumping resonance at
$\Omega_{mod}=n\Omega_L$, producing polarization moments with
$|q|\ge n$, leads to resonance features in the signal
\cite{FMNMOR,Malakyan,BudkerSelective}. Because the PM components
have $|q|$-fold symmetry, the signal that they create in the probe
beam is at the frequency $|q|\Omega_L$ (Fig.\
\ref{fig:AveragedPMs}).
Thus if observations are made at the first harmonic of the
modulation frequency, only the components with $|q|=n$ will
contribute to the signal. For example, $^{87}$Rb has a resonance
related to the $|q|=2$ components of the quadrupole and hexadecapole
moments ($\rho^{(2)}_{\pm2}$ and $\rho^{(4)}_{\pm2}$) appearing at
$\Omega_{mod}=2\Omega_{L}$, and a resonance related to the $|q|=4$
components of the hexadecapole moment ($\rho^{(4)}_{\pm4}$) at
$\Omega_{mod}=4\Omega_{L}$. In $^{85}$Rb an extra resonance appears
at $\Omega_{mod}=6\Omega_{L}$, which it is related to the $|q|=6$
components of the hexacontatetrapole ($\kappa=6$) moment
($\rho^{(6)}_{\pm6}$).

The appearance of these narrow resonances at high magnetic fields
allows NMOR-based magnetometry with no trade-off between dynamic
range and sensitivity. In addition, the separated pump and probe
technique used in this work enables independent investigation of
creation and detection of the PMs.

\section{Experimental setup \label{sec:setup}}

\begin{figure}
    \includegraphics{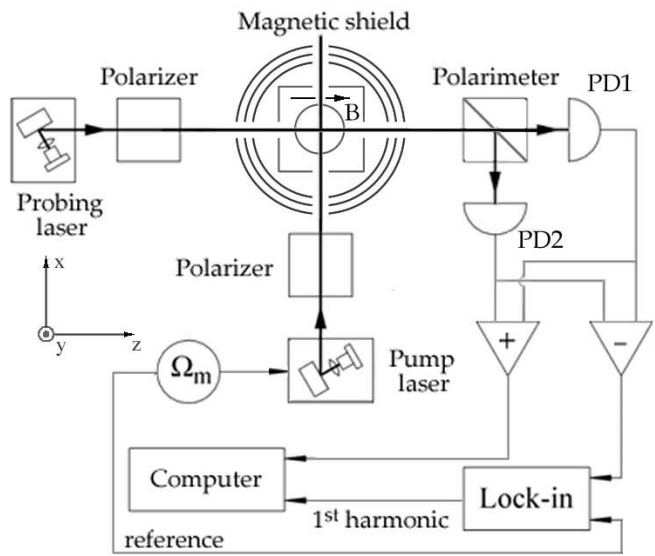}
    \caption{Experimental setup. PD1 and PD2 are photodiodes.}
    \label{fig:setup}
\end{figure}

The experiment is an extension of the previous experiments
\cite{FMNMOR,BudkerSelective,Malakyan} in which FM NMOR was studied
with a single light beam. The scheme of the present experiment is
shown in Fig.\ \ref{fig:setup}.
The atomic vapor cell is placed in a four-layer magnetic shield. The
shield provides passive attenuation of dc fields to a level of 1
part in 10$^6$ \cite{YashchukShield}. Magnetic coils mounted inside
the shield are used for compensation of residual magnetic fields and
their first-order gradients. They are also used to produce a bias
magnetic field along the $z$-axis.

The separated pump and probe beams, each 2 mm in diameter, interact
with an isotopically enriched sample of $^{85}$Rb or $^{87}$Rb atoms
contained in a buffer-gas-free, paraffin-coated, 10 cm diameter
spherical cells. The coating of the cell walls allows the lifetime
of atomic PMs in the ground state in the absence of light to be up
to \linebreak[4] 500 ms. The external-cavity diode laser (pump
laser) produces light at 795 nm for the rubidium D1 line
($^2$S$_{1/2} \rightarrow^2$P$_{1/2}$). In most cases, for $^{87}$Rb
measurements, the central frequency of the laser is tuned to the
low-frequency wing of the $F=2\rightarrow F'=1$ transition (see
$^{87}$Rb spectrum of the D1 line in Fig.\
\ref{fig:PumpSpectrum}(c)), while for $^{85}$Rb measurements it is
tuned to the center of the $F=3\rightarrow F'$ transition group (see
$^{85}$Rb spectrum of the D1 line in Fig.\
\ref{fig:PumpSpectrum85Rb}(b)). In both cases the laser central
frequency is stabilized with a dichroic atomic vapor laser lock
\cite{DAVLL,BudkerDAVLL} modified for operation with frequency
modulated light. Synchronous pumping of atoms is achieved with
frequency modulated pump-light. The modulation frequency ranged from
70 Hz up to 1 kHz with 300 MHz (peak to peak) modulation depth. The
unmodulated probe laser generates light at 780 nm for the rubidium
D2 line ($^2$S$_{1/2} \rightarrow ^2$P$_{3/2}$). The frequency of
the probe laser is stabilized and tuned, in most cases for the
$^{87}$Rb measurements, to the center of the $F=2\rightarrow F''$
transition group (see $^{87}$Rb spectrum of the D2 line in Fig.\
\ref{fig:ProbeSpectrum}(c)) and, for $^{85}$Rb measurements, to the
center of the $F=3\rightarrow F''$ transition group. The use of two
light beams interacting with atoms via different excited states (D1
and D2 transitions) ensures, in addition to the small crossing area
of the pump and probe beams inside a shield, that the optical
rotation is only due to the polarization in the ground state
\footnote{The rotation of the polarization plane due to the
Bennett-structure effect (see, for example, Ref.\
\cite{Bud2002Bennett}) is negligible in our experiment.}.

The light beams are perpendicular; the pump beam propagates along
$x$-axis and the probe beam propagates along $z$-axis, parallel to
the magnetic field. Both beams are linearly polarized along $y$-axis
before entering a cell.

Upon passing through the cell the polarization of the probe beam is
modulated due to the precessing anisotropy of the medium. The
polarization is analyzed using a balanced polarimeter. In the
polarimeter two photodiodes detect transmission through a polarizing
beam-splitter at $45^\circ$ to the incident light beam polarization
in the $x$-$y$ plane. The amplitude of the photodiode difference
signal is detected with a lock-in amplifier at the first harmonic of
$\Omega_{mod}$. The FM NMOR resonances were recorded by scanning
$\Omega_{mod}$.

\section{Results and Analysis}
\label{sec:results}

In this section the experimental results and their analysis are
presented. The typical two-beam FM NMOR signal is discussed and
compared with the single-beam FM NMOR signal. We then discuss the
dependences of the two-beam signals on the pump- and probe-beam
parameters, to understand the detailed mechanisms of creation and
detection of atomic multipoles.

\begin{figure}
    \includegraphics{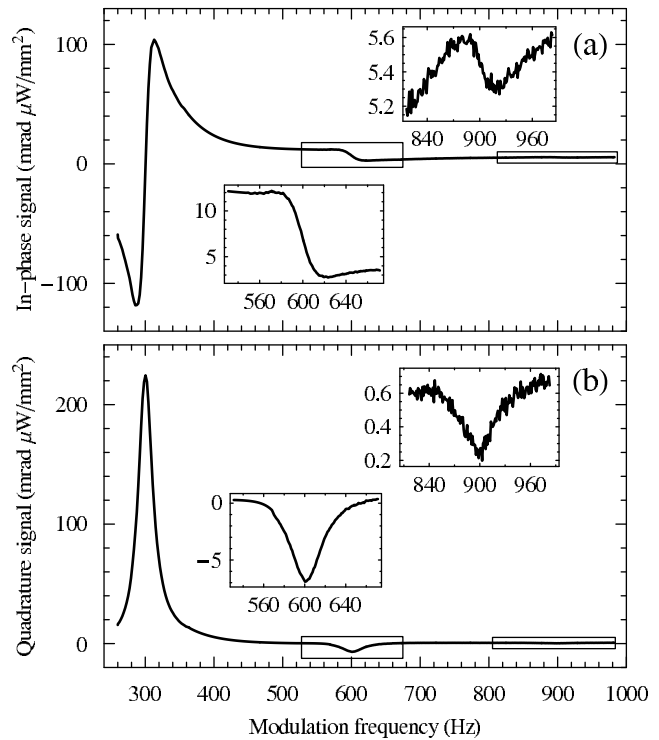}
    \caption{The two-beam FM NMOR in-phase (a) and quadrature (b)
    signals in the modulation-frequency domain with fixed magnetic
    field ($\Omega_L\approx2\pi\times 150$ Hz). Signals at
    $\approx\,$300 Hz, $\approx\,$600 Hz, and $\approx\,$900Hz are related to the
    two-fold symmetric multipoles (mostly quadrupole), four-fold
    symmetric multipoles (mostly hexadecapole) and six-fold symetric
    (hexacontatetrapole) moment, respectively. The insets show
    magnification of the hexadecapole and hexacontaterapole resonances. The signals were
    recorded in $^{85}$Rb with $I_{pump}=17$ $\mu$W/mm$^2$ and
    $I_{probe}=20$ $\mu$W/mm$^2$. Pump laser was tuned to the center
    of the $F=3\rightarrow F'$ transition group and probe laser to
    the high-frequency wing of the $F=3\rightarrow F''$ transition group. The units of the signal
    correspond to the rotation of the polarization plane of the
    probe beam multiplied by the probe-beam intensity. In order to
    calculate the angle of the rotation the recorded signal should
    be divided by the probe-light intensity.}
    \label{fig:FMNMORSignal}
\end{figure}

A representative recording of the two-beam FM NMOR signal in the
modulation-frequency domain measured in $^{85}$Rb is presented in
Fig.\ \ref{fig:FMNMORSignal}.
Three resonances were observed centered at twice ($\approx\,$300
Hz), four times ($\approx\,$600 Hz), and six times ($\approx\,$900
Hz) the Larmor frequency ($\Omega_L\approx2\pi\times150$ Hz). As
discussed in Section \ref{sec:metodology}, these resonances are
related to the two-fold symmetric multipoles (mostly quadrupole
$\rho^{(2)}_{\pm2}$ but also with some hexadecapole
$\rho^{(4)}_{\pm2}$ and hexacontatetrapole $\rho^{(6)}_{\pm2}$
admixture), four-fold symmetric hexadecapole and hexacontatetrapole
moments, $\rho^{(4)}_{\pm4}$ and $\rho^{(6)}_{\pm4}$, respectively
and pure six-fold symmetric hexacontatetrapole $\rho^{(6)}_{\pm6}$.
According to the authors' best knowledge, signal observed at
six-times the Larmor frequency (hexacontatetrapole resonance) is the
first observation of an effect related solely to the $\Delta m=6$
coherences. A detailed analysis of selective creation and detection
of this type of coherences is given in Section \ref{sec:results} C.

\begin{figure}
    \includegraphics{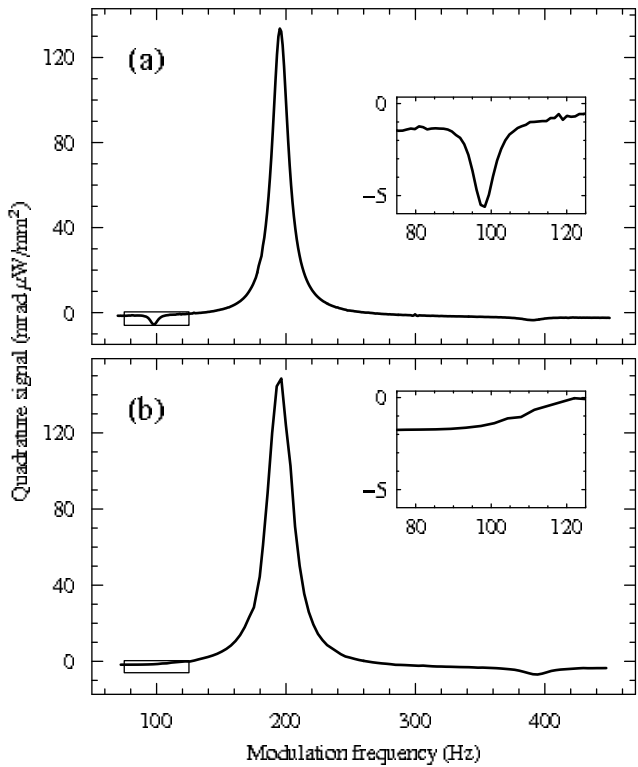}
    \caption{Single-beam (a) and two-beam (b) FM NMOR signals
    measured in $^{87}$Rb. The upper inset shows a magnification of
    the resonance ($\Omega_{mod}\approx2\pi\times 97.5$ Hz)
    appearing due to the modulation of the probe light in
    single-beam FM NMOR. The lower inset shows the absence of that
    signal in the two-beam arrangement. The pump-beam central frequency and
    probe-beam frequency were tuned to the low-frequency wing of the $F=2\rightarrow
    F'=1$ transition of the D1 line and the center of the
    $F=2\rightarrow F''$ transition group of the D2 line,
    respectively, in the two-beam experiment. In the single-beam FM
    NMOR experiment light was tuned to the low-frequency wing of the
    D1 line. The intensities of the pump and probe in the two-beam
    experiment were the same ($I_{pump}=I_{probe}=20$
    $\mu$W/mm$^2$), half of the intensity of the incident light in
    the single-beam experiment ($I=40$ $\mu$W/mm$^2$).}
    \label{fig:OneTwo}
\end{figure}

A comparison of single-beam and two-beam resonances is given in
Fig.\ \ref{fig:OneTwo}.
In the single-beam FM NMOR experiment
\cite{FMNMOR,BudkerSelective,Malakyan} the medium is pumped and
probed with the same, frequency-modulated laser beam. Because in
this case the probe light is modulated, additional resonances occur
that do not appear in the two-beam case (see insets in Fig. 4). At
$\Omega_{mod}=2\Omega_L$ the quarupole is efficiently pumped (this
is a resonance which is part of the an $\Omega_{mod}=2\Omega_L$
family of resonances for quadrupole), but there is no signal
observed at the first harmonic for unmodulated probe light as the
polarization is modulated by the medium at 2$\Omega_L$, i.e., at the
second harmonic of $\Omega_{mod}$. However, when the probe is
modulated, this leads to the first-harmonic signal as observed in
the single-beam experiment (Fig.\ \ref{fig:OneTwo}(a)). Thus, unlike
the two-beam experiment with unmodulated probe, a resonance appears
at $\Omega_{mod}=\Omega_L$ (compare the resonance at
$\Omega_{mod}\approx2\pi\times 97.5$ Hz in Fig.\ \ref{fig:OneTwo}(a)
with the absence of one in Fig.\ \ref{fig:OneTwo}(b)).

Below we present the dependence of FM NMOR resonances' amplitudes
and widths as a function of pump- and probe-light parameters. The
amplitudes and widths of the FM NMOR resonances were determined by
fitting the dispersive-like Lorentzians
($Ax/(x^2+\delta\Omega_{mod}^2$) to the in-phase component and
absorptive-like Lorentzians ($A/(x^2+\delta\Omega_{mod}^2$) to the
quadrature component of the recorded signals. Within our
experimental conditions the amplitudes and widths determined in both
components of the signals are the same.

The discussion in the two next subsections refers to data taken with
$^{87}$Rb. The $^{87}$Rb data obtained as a function of pump- and
probe-beam parameter are considered separately in order to
facilitates the comparison between creation and detection of the
atomic PMs. In Section\ \ref{sec:hexacontatetra} the $^{85}$Rb data
is discussed.

\subsection{Creation of atomic PMs}

\begin{figure}
    \includegraphics{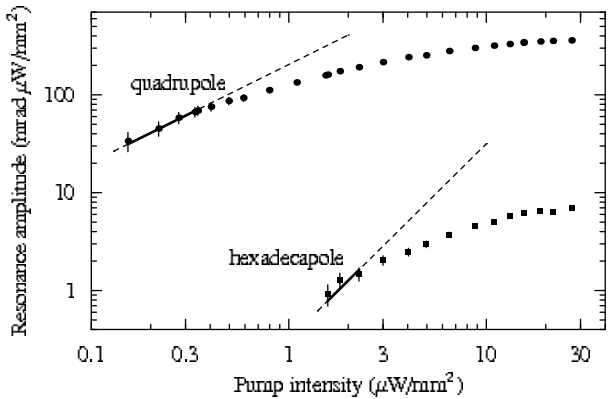}
    \caption{The amplitude of the quadrupole and hexadecapole
    signals vs.\ pump-beam intensity for $^{87}$Rb. The
    quadrupole signal grows linearly with pump-light intensity at low
    intensities, and then levels off due to saturation. Under these
    experimental conditions the hexadecapole signal was not large
    enough to be observed at intensities much below saturation.
    The two dashed lines show linear and quadratic slopes in
    pump intensity. The solid portions show the regions in which the experimental points obey linear and
    quadratic dependences, respectively. The intensity of the probe,
    tuned to the center of the $F=2\rightarrow F''$ transition
    group, was 1.6 $\mu$W/mm$^2$ and the pump-beam central frequency
    was tuned to the low-frequency wing of the $F=2\rightarrow F'=1$
    transition group.}
    \label{fig:PumpAmplitude}
\end{figure}

The amplitudes of the quadrupole and hexadecapole FM NMOR resonances
are plotted vs.\ pump-light intensity in Fig.\
\ref{fig:PumpAmplitude}.
The probe beam was attenuated to 1.6 $\mu$W/mm$^2$ to simplify the
interpretation of the results. However, the drawback of choosing
such probe-light intensity is that at low pump-beam intensities only
the quadrupole signal could be observed. The hexadecapole signal was
measured at pump intensities above 5 $\mu$W/mm$^2$.

At low intensities, the quadrupole amplitude scales linearly with
pump-light intensity. This is because only a one-photon interaction
is needed for the creation of the quadrupole moment. However, as the
pump-beam intensity increases the curve tends to grow slower than
linear with $I_{pump}$ due to higher-order effects. We will refer to
these generally as saturation effects, although they may include
more complex processes such as alignment-to-orientation conversion
and related phenomena (see, for example, Ref.\ \cite{Bud2000AOC}).
Although the lack of low pump-intensity data for the hexadecapole
moment preclude reliable determination of the intensity dependence
in the low-intensity limit, the data are not inconsistent with the
expected quadratic behavior.

As can be seen in Fig.\ \ref{fig:PumpAmplitude} the amplitude of the
quadrupole and hexadecapole signals differ by at least two orders of
magnitude. As pointed out by M. Auzinsh, it may be possible to
generate the hexadecapole moment more efficiently by pumping at
$2\Omega_L$ rather than $4\Omega_L$. This is because the quadrupole
moment must be present in order to pump the hexadecapole moment, but
the transverse quadrupole moment is averaged out when
$\Omega_{mod}=4\Omega_L$, as discussed in Section\
\ref{sec:metodology}, but will be not averaged out at
$\Omega_{mod}=2\Omega_L$. The signal due to the hexadecapole moment
could then be separated from the quadrupole signal, in a two-beam
arrangement, by detecting the signal at $4\Omega_L$ in an
unmodulated probe. A detailed discussion of this point will be given
elsewhere \cite{HexAt2Larmor}.

\begin{figure}
    \includegraphics{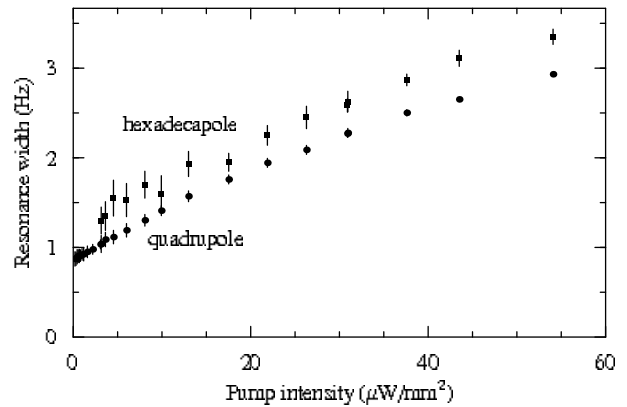}
    \caption{The widths of quadrupole and hexadecapole resonances
    as a function of pump-beam intensity. The intensity of the probe is 1.6
    $\mu$W/mm$^2$. Both beams were tuned the the same transitions as
    for the amplitude pump-intensity dependences (Fig.\ \ref{fig:PumpAmplitude}).}
    \label{fig:PumpWidth}
\end{figure}

The widths of the quadrupole and hexadecapole FM NMOR resonances are
plotted vs.\ pump intensity in Fig.\ \ref{fig:PumpWidth}.
The experimental conditions are similar to those used for the
measurement of the amplitude dependences. The low intensity of the
probe (1.6 $\mu$W/mm$^2$) ensures that the power broadening of the
signal due to the probe beam is small \footnote{While the width
dependences of both PMs shown in Fig. 7 look similar, the detailed
mechanisms responsible for the two broadenings are di®erent. This is
due to the fact, that higher-order multipoles are created from the
lower-rank PMs by interaction with additional photons. In
consequence, just as more photons are needed to cre- ate higher-rank
PMs, more photons contribute also to their saturation. Although the
exact functional depen- dences of the widths of the FM NMOR
resonances vs. light power is unknown we approximate them with
linear dependences in low-light limit.}. The widths of the
resonances doubly extrapolated to zero pump- and probe-light
intensity are $\delta\Omega_{mod}^{(2)}=2\pi\times 0.85(5)$ Hz for
the quadrupole moment and $\delta\Omega_{mod}^{(4)}=2\pi\times
1.2(1)$ Hz for the hexadecapole moment, respectively. Since these
widths are proportional to the relaxation rate of the respective
multipoles, this implies that the quadrupole moment decays more
slowly than the hexadecapole moment. The experimental ratio between
the widths of the quadrupole and hexadecapole resonances doubly
extrapolated to zero intensities of pump and probe beams is 0.7(1).
This is in approximate agreement with the theoretical prediction of
the electron randomization model \cite{Happer72} expected to be the
main mechanism of relaxation, which gives the ratio as 9/16
\footnote{In Ref.\ \cite{BudkerSelective}, the expected ratio of the
relaxation was erroneously quoted as 3/8 (instead of 9/16). This
change of the numerical value does not affect the qualitative
conclusions reached in that paper. The experimental results in Ref.\
\cite{BudkerSelective} corresponds to
$\delta\Omega_{mod}^{(2)}/\delta\Omega_{mod}^{(4)}=0.47(4)$.}. Other
mechanisms of relaxation \cite{RelaxationInTheDark} that are the
same for both multipoles may contribute to the deviation between the
experimental and theoretical ratio.

\begin{figure}
    \includegraphics{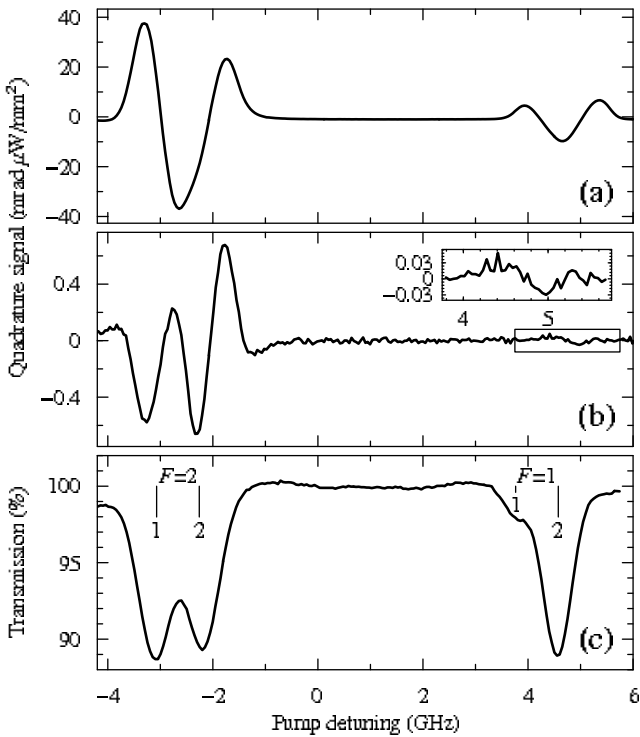}
    \caption{The $^{87}$Rb FM NMOR spectra as a function of the
    pump-light central frequency for the quadrupole (a) and hexadecapole (b) resonances
    along with the transmission spectrum (c). Note the different
    spectral dependences of the multipoles. The signal components
    observed at the $F=1\rightarrow F'$ transition result from the
    coherence transfer from the excited state $F'$ to the $F=2$
    ground state via spontaneous emission (The inset shows the
    magnification of the signal related to the transfer of the
    $\Delta m=4$ coherences). The vertical lines in
    plot (c) show the relative transition strengths of the
    $F\rightarrow F'$ hyperfine transitions, where the number below
    each line indicates the value of $F'$. The probe laser was tuned
    to the center of the $F=2\rightarrow F''$ transition group and
    its intensity was set to be 3.2 $\mu$W/mm$^2$. The pump
    intensity was 59 $\mu$W/mm$^2$. The absorption spectrum
    (c) was recorded at low light intensity (2.6
    $\mu$W/mm$^2$).} \label{fig:PumpSpectrum}
\end{figure}

The quadrupole and hexadecapole spectra measured as a function of
the pump-light central frequency are shown in Fig.\
\ref{fig:PumpSpectrum}.
The probe laser was locked to the center of the $F=2\rightarrow F''$
transition group while the pump-beam central frequency was scanned
over a range covering all hyperfine components of the D1 line. The
signal in Fig.\ \ref{fig:PumpSpectrum}(a) related to the quadrupole
moment was recorded with the modulation frequency fixed at 194 Hz,
which was the frequency that produced the maximal rotation in the
quadrature component of the signal. For the much smaller
hexadecapole signal, we subtracted from the spectrum recorded at 388
Hz (the frequency producing the maximum quadrature hexadecapole
signal) the spectrum measured for a modulation frequency detuned a
few resonance widths away. This is done to remove the effect of the
off-resonant background rotation on the recorded signal. The signal
related only to the hexadecapole moment is shown in Fig.\
\ref{fig:PumpSpectrum}(b).

Because the $F=2$ ground state can support both the quadrupole and
hexadecapole moments, we would expect that signals due to both these
moments would be observed for the pump beam tuned to the
$F=2\rightarrow F'$ transition group, as Fig.\
\ref{fig:PumpSpectrum} shows. However, the fact that signals related
to both PMs are also observed for pump light tuned to the
$F=1\rightarrow F'$ transition group is less intuitive. The
polarization of the $F=1$ ground state cannot be detected with the
probe light tuned to $F=2\rightarrow F''$ which is only sensitive to
the polarization of the $F=2$ state. Furthermore, polarization
cannot be transferred between the ground states via spin exchange
because the two ground states experience rapid Larmor precession in
opposite directions. Thus, the multipoles in the $F=2$ state can
only be generated indirectly in the following process. The pump beam
tuned to the $F=1\rightarrow F'$ transition group creates multipoles
in the excited states. The $F'=2$ state can support both the
quadrupole and hexadecapole moments. The excited-state multipoles
are then transferred to the $F=2$ ground state via spontaneous
emission. Such PM transfer has been observed previously for the case
of the quadrupole moment \cite{Auzinsh89,Ale93}. However, this is
the first observation of the transfer of higher-order coherences.

\subsection{Signature of the PMs}

\begin{figure}
    \includegraphics{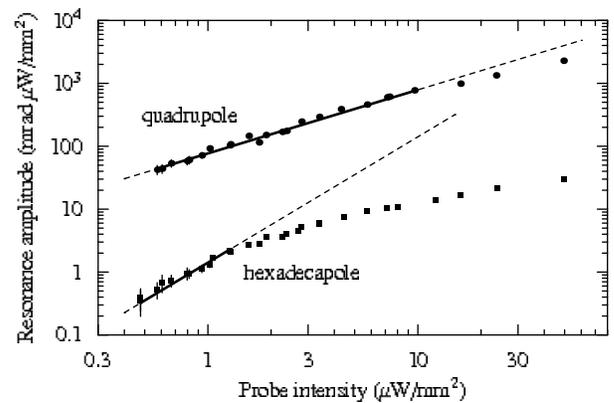}
    \caption{The amplitudes of the quadrupole and hexadecapole
    signals vs.\ probe-light intensity at high pump-light intensity
    (70 $\mu$W/mm$^2$). The solid lines indicate the region in which
    the quadrupole has approximately linear and the hexadecapole
    approximately quadratic dependence on probe-beam intensity. The
    signals were recorded with the central frequency of the pump laser tuned to the low-frequency
    wing of the $F=2\rightarrow F'=1$ transition and the
    probe beam tuned to the center of the $F=2\rightarrow F''$
    transition group of $^{87}$Rb.} \label{fig:ProbeAmplitude}
\end{figure}

The amplitudes of the quadrupole and hexadecapole FM NMOR signals
are plotted as a function of probe-light intensity in Fig.\
\ref{fig:ProbeAmplitude}.
The pump and probe beams have the same tuning and modulation
frequencies as in the previous cases. The pump-beam intensity is
high ($I_{pump}=70$ $\mu$W/mm$^2$), enabling observation of the
hexadecapole moment at relatively low probe-beam intensities.

For low probe intensity the amplitudes of the signals corresponding
to both PMs asymptotically obey theoretical predictions: the
quadrupole signal scales linearly and the hexadecapole quadratically
with the probe-light intensity, the same as the pump-intensity
dependence (see Fig.\ \ref{fig:PumpAmplitude} for comparison). This
is expected, as the same number of light-atom interactions is needed
for creation and for detection of a given atomic PM. For both the
quadrupole and hexadecapole signals, the products of the pump and
probe intensity dependences are consistent with the intensity
dependences previously observed in single-beam FM NMOR
\cite{BudkerSelective}.

\begin{figure}
    \includegraphics{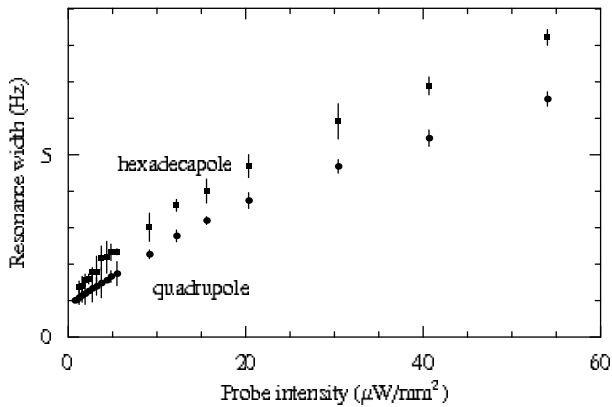}
    \caption{The widths of the FM NMOR signals vs.\ probe-light
    intensity. The pump beam was tuned to the low-frequency wing of the $F=2\rightarrow
    F'=1$ transition and the probe light was tuned near the
    center of the $F=2\rightarrow F''$ transition group of
    $^{87}$Rb. The pump intensity was set to 8 $\mu$W/mm$^2$.}
    \label{fig:ProbeWidth}
\end{figure}

The widths of the quadrupole and hexadecapole resonances as a
function of probe-beam intensity are shown in Fig.\
\ref{fig:ProbeWidth}. To reduce the influence of pump-saturation
effects on the measured signals the pump-light intensity was chosen
to be relatively low (8 $\mu$W/mm$^2$). However, this pump-light
intensity is still too high to allow for a reliable extrapolation to
extract the intrinsic resonance widths.

With increasing probe-light intensity both resonances exhibit power
broadening. The observed broadening is stronger than the broadening
of the resonances measured as a function of pump-beam intensity (for
comparison see Fig.\ \ref{fig:PumpWidth}) which is partly a result
of the fact that the pump beam is modulated and so only periodically
causes broadening and partly is due to different transition dipole
moments of the D1 and D2 lines.

\begin{figure}
    \includegraphics{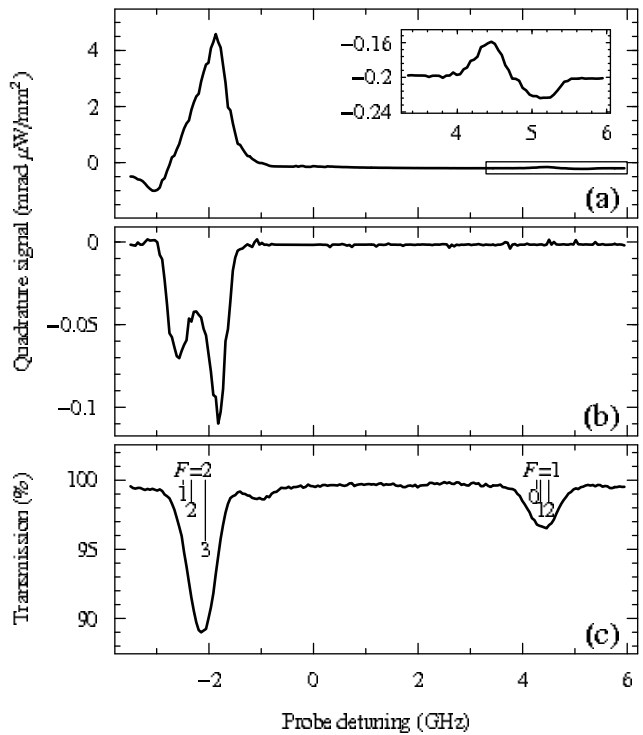}
    \caption{Quadrupole (a) and hexadecapole (b) signal as
    a function of probe-beam frequency; (c) reference
    absorption spectrum of $^{87}$Rb . The inset shows the
    magnification of the quadrupole signal related to the transfer
    of the coherences $\Delta m=2$ from the excited $F'=1$ or $F'=2$ state to the
    ground state $F=2$.
    In order to avoid saturation
    effects due to the pump light, the pump-beam intensity was relatively low
    (3.2 $\mu$W/mm$^2$). This entails the use of a high probe-beam
    intensity (57 $\mu$W/mm$^2$). The pump-beam central frequency was tuned to the
    low-frequency wing of the $F=2\rightarrow F'=1$ transition.}
    \label{fig:ProbeSpectrum}
\end{figure}

The measured spectral dependences of the quadrupole and hexadecapole
resonances as a function of probe-beam frequency are shown in Fig.\
\ref{fig:ProbeSpectrum}.
The probe frequency was scanned over all hyperfine components of the
D2 line. The quadrupole signal is shown in Fig.\
\ref{fig:ProbeSpectrum}(a), and the difference hexadecapole signal
obtained in the same way as for the data presented in Fig.\
\ref{fig:PumpSpectrum} is shown in Fig.\ \ref{fig:ProbeSpectrum}(b).
It is seen that the spectral dependences of the two signals are
significantly different. They also differ from the spectra recorded
vs.\ pump-beam central frequency (see Fig.\ \ref{fig:PumpSpectrum}
for comparison). The difference in frequency dependences for the
pump and the probe beams illustrates the potential offered by the
separated beam technique for optimization of the signal associated
with a given PM.

Strong quadrupole and hexadecapole signals were observed when the
probe light was tuned to the $F=2\rightarrow F''$ transition group.
These PMs were produced directly by the pump light tuned to the
low-frequency wing of the $F=2\rightarrow F'=1$ transition. There
was also very weak quadrupole signal measured when tuned to the
$F=1\rightarrow F''$ transition group, which could only be produced
by spontaneous decay of polarization from the upper state. We have
verified with a calculation that the branching ratios are such that
the generation of the quadrupole moment in the $F=1$ ground state is
strongly suppressed in this case. As expected, hexadecapole signal
is not observed when probing the $F=1$ ground state, which does not
support the hexadecapole moment.

\subsection{Creation and detection of hexacontatetrapole moment}
\label{sec:hexacontatetra}

The discussion in this subsection concerns data taken with
$^{85}$Rb, the $F=3$ ground state of which can support the PM of
rank $\kappa=6$. Due to the fact that the averaged
hexacontatetrapole moment has 6-fold symmetry (see Fig.\
\ref{fig:AveragedPMs}(c)) the FM NMOR signals related to this
multipole are observed at $\Omega_{mod}=6\Omega_L$.

\begin{figure}
    \includegraphics{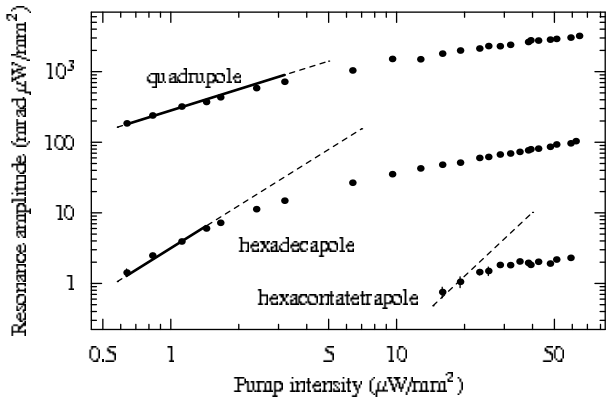}
    \caption{The amplitudes of the FM NMOR resonances related to the
    quadrupole, hexadecapole and hexacontatetrapole moments in
    $^{85}$Rb vs.\ pump-beam intensity. The dashed lines show linear,
    quadratic and cubic slopes while their solid parts indicate the
    region in which recorded dependences obey theoretical
    predictions. Under these experimental conditions (I$_{probe}=36$
    $\mu$W/mm$^2$) the measured amplitude dependence of the
    hexacontatetrapole signal is weaker than predicted cubic
    relation which is a result of saturation. The pump beam central frequency was
    tuned to the center of the $F=3\rightarrow F'$
    transition group while the probe beam was tuned to the high-frequency wing of
    the $F=3\rightarrow F''$ transition group of $^{85}$Rb.}
    \label{fig:PumpAmplitudeRb85}
\end{figure}

\begin{figure}
    \includegraphics{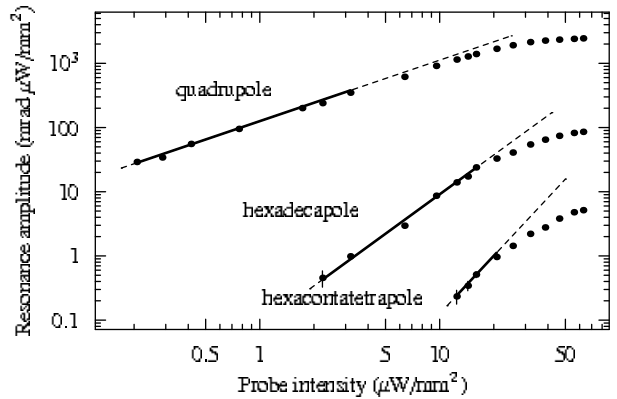}
    \caption{The amplitudes of the FM NMOR signals related to the
    quadrupole, hexadecapole and hexacontatetrapole moments in $^{85}$Rb as a
    function of probe-light intensity. The dashed and solid lines
    have the same meaning as in Fig.\ \ref{fig:PumpAmplitudeRb85}.
    The tuning of the pump beam was the same as in the previous case
    and its intensity was 52 $\mu$W/mm$^2$. The probe beam tuning was slightly
    different than in previous case and it was locked
    to the center of the $F=3\rightarrow F''$
    transition group of $^{85}$Rb. This difference in tuning of the probe-beam
    is a reason why the amplitudes of the FM NMOR signals presented in this
    plot are smaller than those in Fig.\ \ref{fig:PumpAmplitudeRb85}.}
\label{fig:ProbeAmplitudeRb85}
\end{figure}

The quadrupole, hexadecapole, and hexacontatetrapole signal
amplitudes are plotted as a function of pump- and probe-beam
intensities in Figs.\ \ref{fig:PumpAmplitudeRb85} and
\ref{fig:ProbeAmplitudeRb85}, respectively. The pump-intensity
dependence of the FM NMOR resonance related to the
hexacontatetrapole moment is stronger than those of the quadrupole
and hexadecapole resonances and is consistent with the expected
$I_{pump}^3$ dependence at low intensity. (Three photons are needed
for the creation or detection of the hexacontatetrapole moment). The
amplitudes of the quadrupole, hexadecapole and hexacontatetrapole
resonances show, respectively, linear, quadratic, and cubic
dependence on $I_{probe}$ (Fig.\ \ref{fig:ProbeAmplitudeRb85}).


\begin{figure}
    \includegraphics{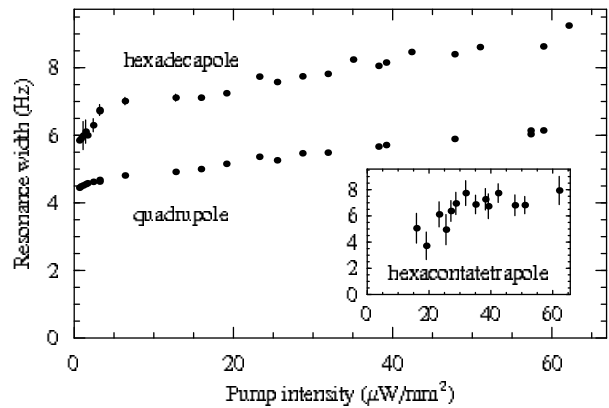}
    \caption{The widths of the $^{85}$Rb quadrupole,
    hexadecapole, and hexacontatetrapole (inset) FM NMOR resonances
    vs.\ pump-light intensity. The central frequency of the pump beam was tuned
    near the center of the $F=3\rightarrow F'$ transition
    group. The probe beam was tuned to the high-frequency wing of the
    $F=3\rightarrow F''$ transition group and its intensity
    was 36 $\mu$W/mm$^2$.} \label{fig:PumpWidthRb85}
\end{figure}

\begin{figure}
    \includegraphics{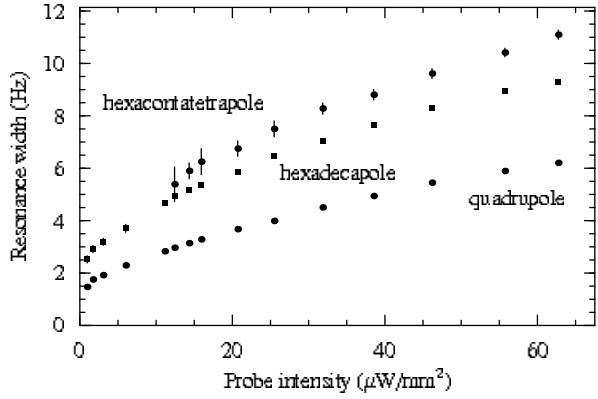}
    \caption{The widths of the FM NMOR signals related to the
    quadrupole, hexadecapole and hexacontatetrapole moments vs.\
    probe-light intensity. The pump beam central frequency was tuned to the center
    of the $F=3\rightarrow F'$ transition group and
    its intensity was 52 $\mu$W/mm$^2$. The probe beam was tuned
    to the center of the $F=3\rightarrow F''$
    transition group.} \label{fig:ProbeWidthRb85}
\end{figure}

The resonance-width dependences for the three multipoles are shown
in Figs.\ \ref{fig:PumpWidthRb85} and \ref{fig:ProbeWidthRb85}.
Although all resonances recorded as a function of pump-beam
intensity (Fig.\ \ref{fig:PumpWidthRb85}) broaden with the light
intensity, the difference in the broadening is significant only in
the first part of the measured dependences.
The difference is due to the different number of photons needed to
create the various multipoles. However, for higher pump-light
intensities the saturation behavior starts to play an important role
and all resonances exhibit similar broadening with the pump-light
intensity i.e. slopes of the width dependences related to a given
multipole are almost the same.

The width dependences of the FM NMOR signals on pump-beam intensity
also show a very interesting feature. For a range of pump
intensities the width of the hexadecapole resonance is broader than
that of the hexaconatetrapole resonance. A more detailed analysis
revealed that the ratio between these widths changes with the tuning
of the pump and probe lasers. For instance, for the probe beam tuned
toward higher frequencies the hexadecapole resonance is broader than
the hexacontatetrapole resonance. The different tuning of the probe
beam is the reason why similar behavior was not observed when
signals were recorded as a function of probe-light intensity (Fig.\
\ref{fig:ProbeWidthRb85}).

\begin{figure}
    \includegraphics{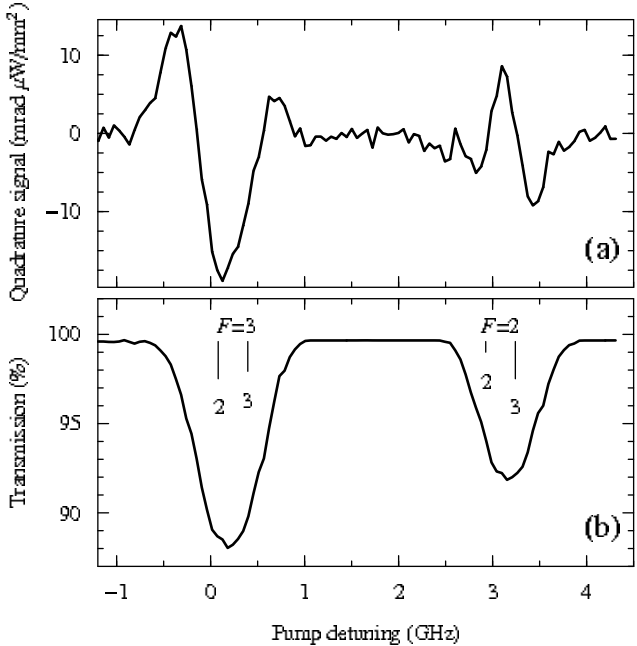}
    \caption{Hexacontatetrapole signal (a) and the
    reference absorption spectrum taken with low light
    intensity (2.5 $\mu$W/mm$^2$) (b) vs. pump-beam
    central frequency recorded with $^{85}$Rb. The strong
    hexacontatetrapole signal is measured at the
    $F=3\rightarrow F'$ transition group which supports
    generation of this multipole in the ground state. The
    hexacontatetrapole signal is also observed for the pump-light
    tuned to the $F=2\rightarrow F'$ transition group which does
    not support creation of the quadrupole in the ground state.
    This signal is due to the transfer of the coherences from
    the excited state $F'=3$ to the ground state $F=3$ via
    spontaneous emission. The probe laser was tuned to the center of
    the $F=3\rightarrow F''$ transition group, while pump-light
    central frequency was scanned over all hyperfine components of $^{85}$Rb
    D1 transition. For the measurements of the signal related to the
    hexacontatetrapole moment the pump-beam intensity was 64 $\mu$W/mm$^2$ and
    probe-beam intensity was 65 $\mu$W/mm$^2$.}
    \label{fig:PumpSpectrum85Rb}
\end{figure}

In Fig.\ \ref{fig:PumpSpectrum85Rb} the hexacontatetrapole signal,
obtained using the method described in Section \ref{sec:results} A,
is shown as a function of pump-light central frequency. The probe
light was tuned to the high-frequency wing of the $F=3\rightarrow
F''$ transition group while the pump beam was scanned over all
$^{85}$Rb hyperfine components of the D1 line.
Under these experimental conditions, a strong hexacontatetrapole
signal was measured at the $F=3\rightarrow F'$ and $F=2\rightarrow
F'$ transition group. The signal recorded at the $F=2\rightarrow F'$
transition group is a result of the hexacontatetrapole-moment
transfer from the $F'=3$ excited state to the $F=3$ ground state via
spontaneous emission, analogously to the situation described in
Section \ref{sec:results} A. The comparable amplitudes of the two
spectral contributions show that the coherence-transfer mechanism
could be very efficient. The amplitude of the hexacontatetrapole
signal recorded at the $F=2\rightarrow F'$ transition group is
comparable with the signal recorded at the $F=3\rightarrow F'$
transition group which directly supports creation of this type of
the coherences in the ground state. This opens potential
applications of this mechanism in the transfer of the Zeeman
coherences between different atomic states.

\begin{figure}
    \includegraphics{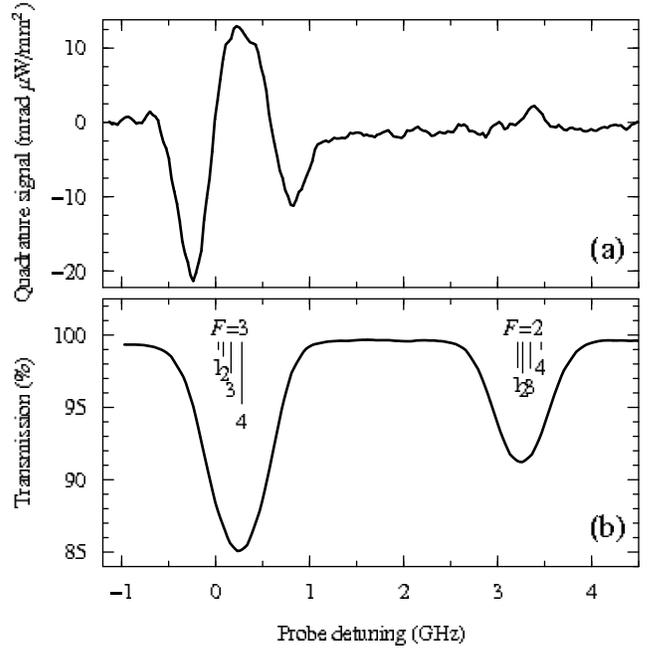}
     \caption{The signal related to the hexacontatetrapole moment
    (a) and the reference spectrum taken for low light intensity
    (b) vs. probe-light frequency. The central frequency of the pump
    beam was tuned to the center of the $F=3\rightarrow F'$ transition group while
    the probe-beam frequency was scanned over all hyperfine components
    of the $^{85}$Rb D2 line. The pump- and probe-beam intensities
    were chosen to be the same as in Fig. \ref{fig:PumpSpectrum85Rb}.}
    \label{fig:ProbeSpectrum85Rb}
\end{figure}

In Fig. \ref{fig:ProbeSpectrum85Rb} the hexacontatetrapole signal is
shown as a function of probe-beam frequency. The spectrum was
recorded for the pump-light central frequency tuned to the
high-frequency wing of the $F=3\rightarrow F'$ transition group. The
probe beam was scanned over all hyperfine components of the
$^{85}$Rb D2 line. The pump- and probe-beam intensities were the
same as for the previous case.
The hexacontatetrapole signal is seen only for the probe beam tuned
to the $F=3\rightarrow F''$ transition group, as expected.

\section{Conclusion}
\label{sec:conclusion}

We have studied the processes of creation and detection of atomic
PMs using the FM NMOR method in a two-beam pump-probe arrangement.
The separated pump and probe beams allow a detailed analysis of
these processes as a function of a number of light-beam parameters
such as intensity, detuning from resonance, and modulation
frequency. The influence of these parameters on the amplitudes and
widths of the FM NMOR resonances has been studied. Working with two
rubidium isotopes, all even-rank PMs supported by the atomic energy
level structure have been detected. It has been shown that signals
due to the quadrupole, hexadecapole, and hexacontatetrapole moments
depend differently on the frequency and intensity of the pump and
probe beams. This may be used for a more complete optimization of
the system than it is possible in a one-beam arrangement. In
particular, a given PM's contribution can be zeroed or maximized,
which may be significant for many applications, especially
magnetometry in the Earth-field range. We have also observed for the
first time the transfer of high-rank moments from the excited to the
ground state via spontaneous emission which may be useful for
creating high-order ground state coherences with minimal light-power
broadening of the FM NMOR resonances.

\begin{acknowledgments}
The authors acknowledge fruitful discussion with Jerzy Zachorowski
and Marcis Auzinsh. S.P. and W.G. would like to thank their
colleagues from UC Berkeley for their hospitality. This work has
been supported by National Science Foundation (grant: 2003-06
INT-0338426), NATO (grant: PST.CLG980362) and by the ONR MURI
program.
\end{acknowledgments}

\end{document}